\documentclass[journal]{IEEEtran}

\pdfoutput=1

\usepackage{amsmath,bm,amssymb,amsfonts,amsthm}
\usepackage[linesnumbered, ruled, vlined]{algorithm2e}
\usepackage{graphicx}
\usepackage{xcolor}
\usepackage{relsize}
\usepackage{textcomp}
\usepackage{listings}

\usepackage{changepage}
\usepackage{hhline}
\usepackage{multirow}
\usepackage{nomencl}
\usepackage{mathtools}
\usepackage{commath}
\usepackage{booktabs}
\usepackage{subfiles}
\usepackage{tabularx}
\usepackage{lscape}
\usepackage{rotating}
\usepackage{amsmath}

\usepackage[normalem]{ulem}
\usepackage[utf8]{inputenc}
\usepackage[hyphens]{url}
\usepackage{lineno}
    \modulolinenumbers[5]
\usepackage[caption=false]{subfig}
\usepackage{enumitem}
    \setlist{nolistsep}
\usepackage{gensymb}
\usepackage[hidelinks]{hyperref}
\usepackage{makecell}
\usepackage{multicol}
\usepackage{fancyhdr}

\usepackage{fancyhdr}
    \fancypagestyle{firstpage}{  \fancyfoot[C]{\textbf{U.S. Government work not protected by U.S. copyright}}}
\usepackage{scalerel}
\usepackage{tikz}
\usetikzlibrary{svg.path}
\definecolor{orcidlogocol}{HTML}{A6CE39}
\tikzset{
  orcidlogo/.pic={
    \fill[orcidlogocol] svg{M256,128c0,70.7-57.3,128-128,128C57.3,256,0,198.7,0,128C0,57.3,57.3,0,128,0C198.7,0,256,57.3,256,128z};
    \fill[white] svg{M86.3,186.2H70.9V79.1h15.4v48.4V186.2z}
                 svg{M108.9,79.1h41.6c39.6,0,57,28.3,57,53.6c0,27.5-21.5,53.6-56.8,53.6h-41.8V79.1z M124.3,172.4h24.5c34.9,0,42.9-26.5,42.9-39.7c0-21.5-13.7-39.7-43.7-39.7h-23.7V172.4z}
                 svg{M88.7,56.8c0,5.5-4.5,10.1-10.1,10.1c-5.6,0-10.1-4.6-10.1-10.1c0-5.6,4.5-10.1,10.1-10.1C84.2,46.7,88.7,51.3,88.7,56.8z};
  }
}
\newcommand\orcidicon[1]{\href{https://orcid.org/#1}{\mbox{\scalerel*{
\begin{tikzpicture}[yscale=-1,transform shape]
\pic{orcidlogo};
\end{tikzpicture}
}{|}}}}

\usepackage[noadjust]{cite}

\begin{document}

\title{\huge Dynamic State Estimation for Load Bus Protection \\ on Inverter-Interfaced Microgrids}

\author{
    Arthur~K.~Barnes $^{1}$\orcidicon{0000-0001-9718-3197},
    Adam~Mate $^{1}$\orcidicon{0000-0002-5628-6509},
    Jean~Marie~V.~Bikorimana $^{2}$, 
    and Ricardo~J.~Castillo $^{3}$ 
    \vspace{-0.25in}

\thanks{Manuscript submitted:~Mar.~18,~2022. 
Current version:~Jun.~3,~2022.
}

\thanks{$^{1}$ The authors are with the Advanced Network Science Initiative at Los Alamos National Laboratory, Los Alamos, NM 87545 USA. \\ Email: abarnes@lanl.gov, amate@lanl.gov.}

\thanks{$^{2}$ The author is with the College of Science and Technologies at the University of Rwanda, Kigali, Rwanda. \\ Email: jbikorimana27@gmail.com.}


\thanks{$^{3}$ The author is with the Energy Security and Resilience Center at National Renewable Energy Laboratory, Golden, CO 80401 USA. \\ Email: ricardo.castillo@nrel.gov.}

\thanks{LA-UR-22-22686. Approved for public release; distribution is unlimited.}

}

\maketitle
\thispagestyle{fancy}
\thispagestyle{firstpage}


\begin{abstract}
Inverter-interfaced microgrids results in challenges when designing protection systems.
Traditional time-overcurrent, admittance, and differential protection methods are unsuitable on account of lack of fault current, excessively short lines, or a prohibitive number of protective devices needing to be installed. Current practice is to force all inverters to shut down during fault conditions, weakening resilience and reducing reliability.
Dynamic state estimation (DSE), which has been explored for both line protection and load bus protection before, is a potential solution to these challenges to create widely utilizable, highly reliable protection systems. However, it has only been tested for load protection with ideal voltage sources, which do not capture the short-circuit behavior of inverter-interfaced generation, notably low fault current and unbalanced output voltage.
This paper aims to extend the state-of-the-art on DSE load protection: the performance of DSE during short-circuit conditions with a grid-forming inverter with current-limiting behavior during fault conditions is investigated.
\end{abstract}
\vspace{-0.05in}
\begin{IEEEkeywords}
power system operation,
microgrid,
distribution network,
protection,
dynamic state estimation.
\end{IEEEkeywords}

\section{Introduction} \label{sec:introduction}
\indent

Proper protection of microgrids -- including small communities, critical healthcare infrastructures, and government facilities -- is increasingly important with the occurring decentralization of the bulk energy system.
Ensuring protection coordination of inverter-interfaced microgrids present a particular challenge due to their unique characteristics: the limited, extremely low fault currents blind conventional protection schemes, rendering them useless \cite{mcdermott_protective_2018, barnes21-pmsp}; when transitioning from grid-connected to islanded operation mode, fault current levels are greatly vary and can flow bidirectionally, making fault detection difficult \cite{tumilty_approaches_2006, liu2020}. The need to adopt different control schemes, makes the fault current analysis and estimation more complex \cite{dehghanpour_2021}.
These challenges limit the size of microgrids in terms of the number of loads served while providing reliable electrical power.

With the introduction and deployment of intelligent electronic devices (IEDs) -- e.g., remote terminal units, phasor measurement units, meters -- over the past decades, high accuracy GPS-synchronized measurements enabled the development of ``setting-less protection'', a.k.a dynamic state estimation (DSE) for both transmission and distribution networks \cite{xie2020, xie2019, meliopoulos2017}.
For microgrid protection, DSE offers a reduced likelihood of misoperation, particularly in the case of protected devices with nonlinear characteristics, and is useful in cases where distance protection performs poorly \cite{meliopoulos2017, barnes21-dse}.
DSE has been investigated as a solution for line protection that can operate under low fault current conditions, though it requires voltage and current measurements on all terminals of the protected elements \cite{liu2015b, vasios2018}.
It has been used to develop centralized protection schemes \cite{choi2017, albinali2017} and to detect hidden failures in substation protection systems \cite{albinali2017b}.
More recently, it has been proposed for protection of load buses and downstream radial sections of a microgrid \cite{liu2018, barnes21-dse, dehghanpour2021}.

Existing work studying DSE for load bus protection only considered operation with ideal voltage sources. However, inverter-interfaced generation will transition to current limiting mode during faults. This means that fault current magnitudes are not significantly higher than current magnitudes at rated power. Additionally, current-limiting on a per-phase basis will result in unbalanced voltage being present on the system. This raises the concern as to if DSE will work under such conditions.
This paper demonstrates the applicability of DSE for the protection of load buses of inverter-interfaced microgrids. Load buses often present a large number of components to protect in a microgrid, therefore their adequate protection is essential to achieve network resilience and grid reliability.
Instead of using ideal power sources like in \cite{barnes21-dse}, non-ideal voltage sources -- specifically grid-forming inverters with current-limiting behavior during fault conditions, introduced in \cite{barnes2021implementing} -- supply the loads. Additionally, downstream radial portions of microgrids are modeled as lumped loads.
Considering this, the performance of DSE during short-circuit conditions is investigated.

\section{Methodology} \label{sec:methodology}
\indent

\begin{figure*}[!htbp]
\centering
\subfloat[Single-phase RL series load]{\includegraphics[width=1.35in]{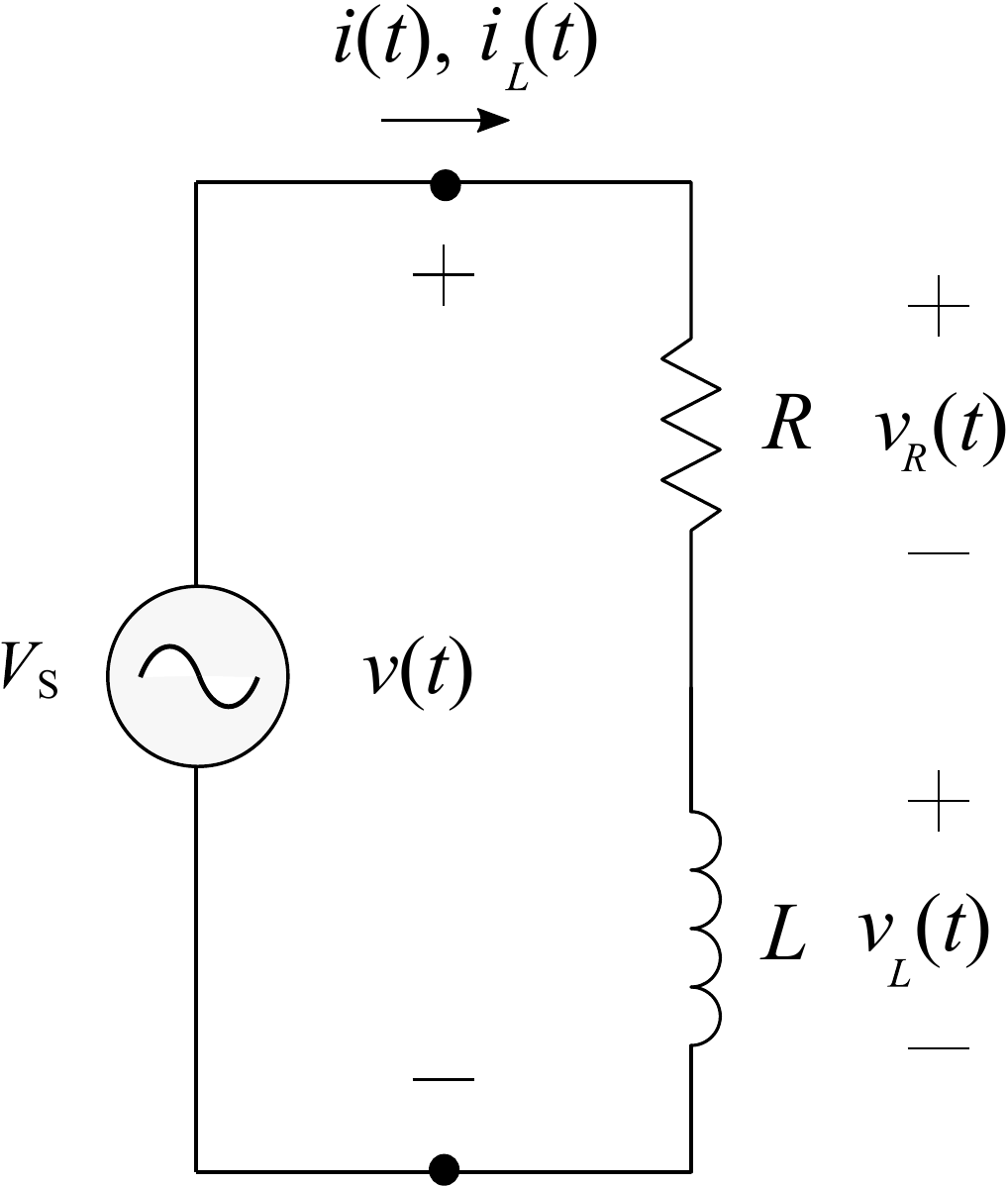}
\label{fig:rl-dynamic}}
\hfil
\subfloat[Grounded-wye-connected RL load]{\includegraphics[width=2.95in]{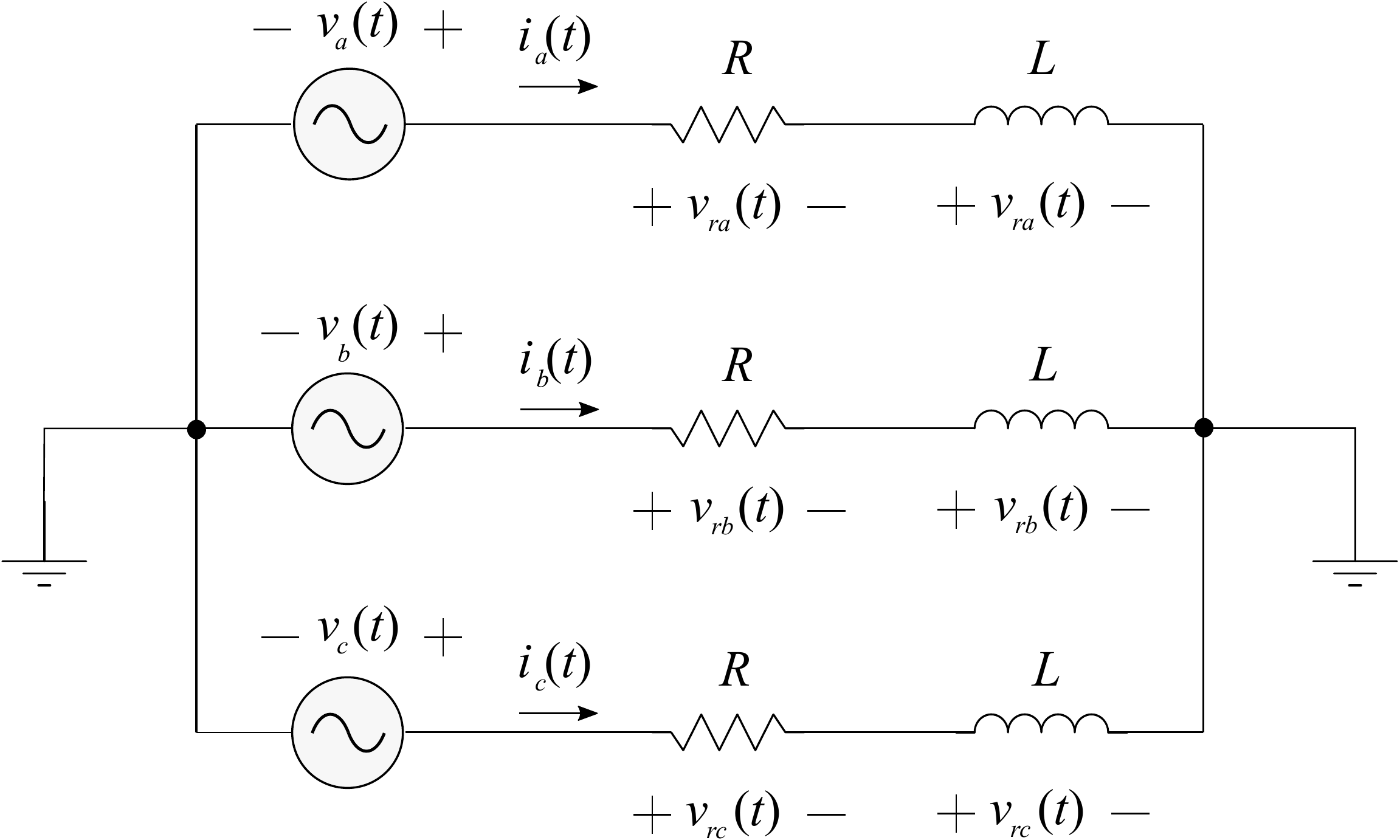}
\label{fig:gwye-no-fault-dynamic}}

\subfloat[Grounded-wye-connected RL load with a line-ground fault]{\includegraphics[width=2.95in]{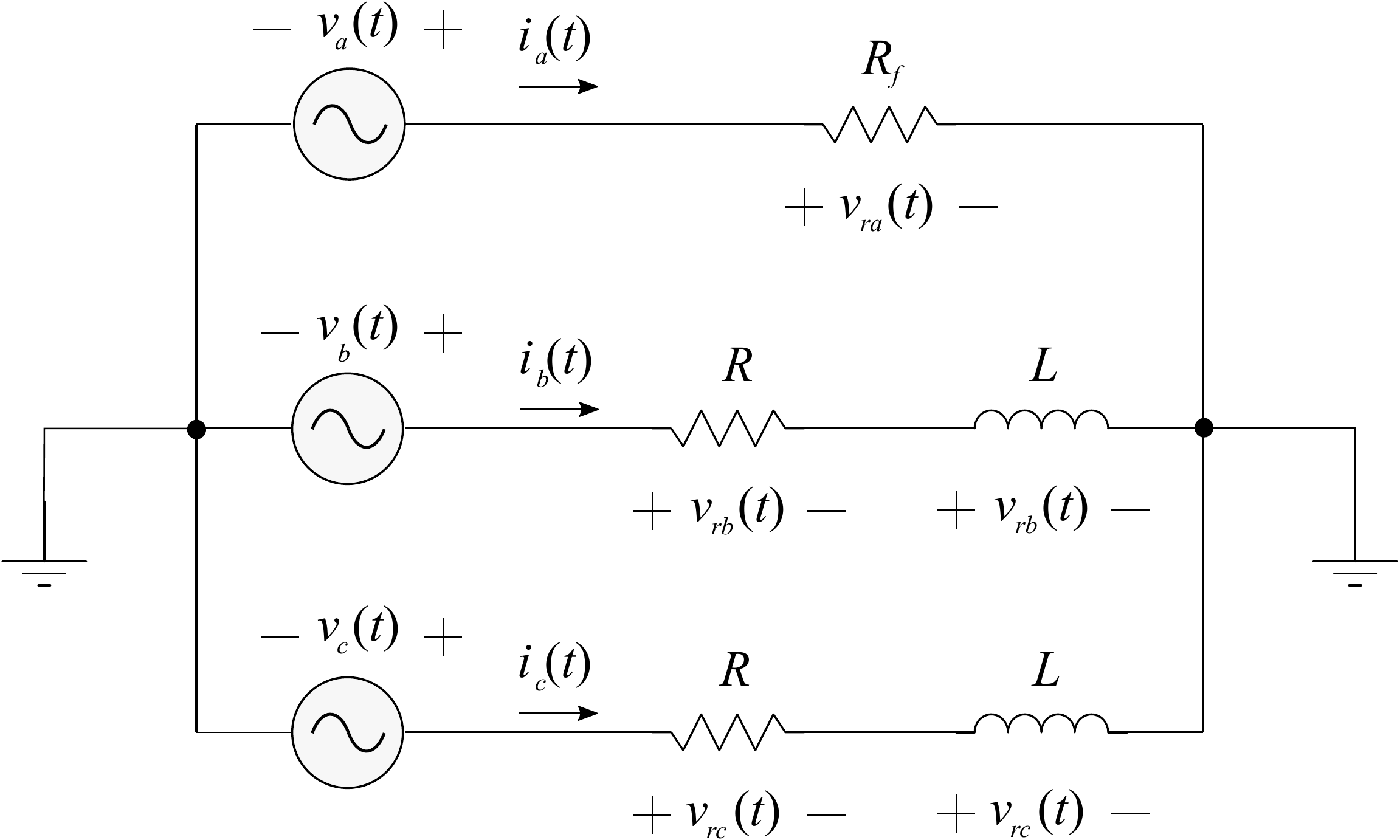}
\label{fig:gwye-lg-fault-r-only-dynamic}}
\hfil
\subfloat[Grounded-wye-connected RL load with a line-line fault]{\includegraphics[width=3.35in]{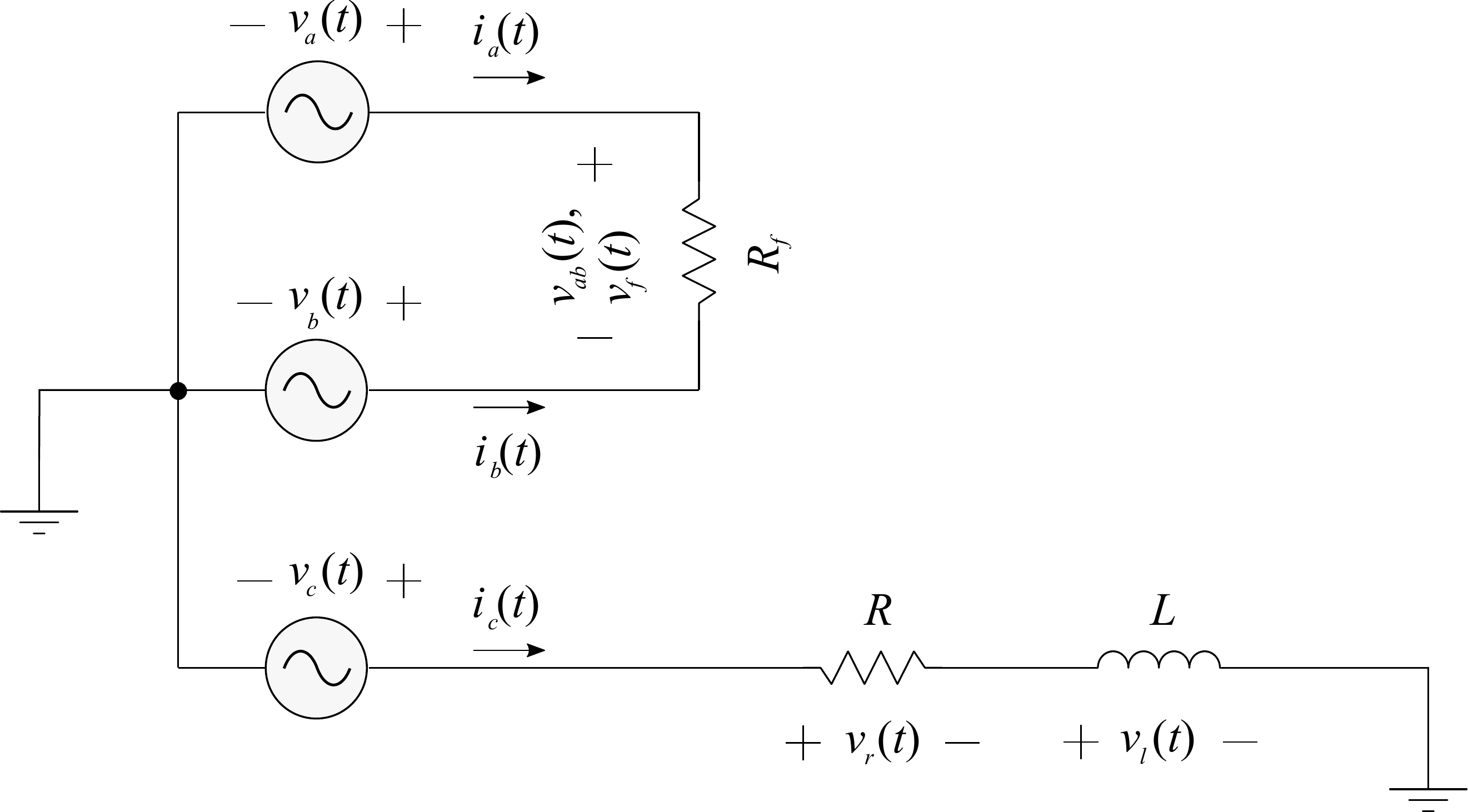}
\label{fig:gwye-ll-fault-r-only-dynamic}}

\subfloat[Delta-connected RL load]{\includegraphics[width=2.0in]{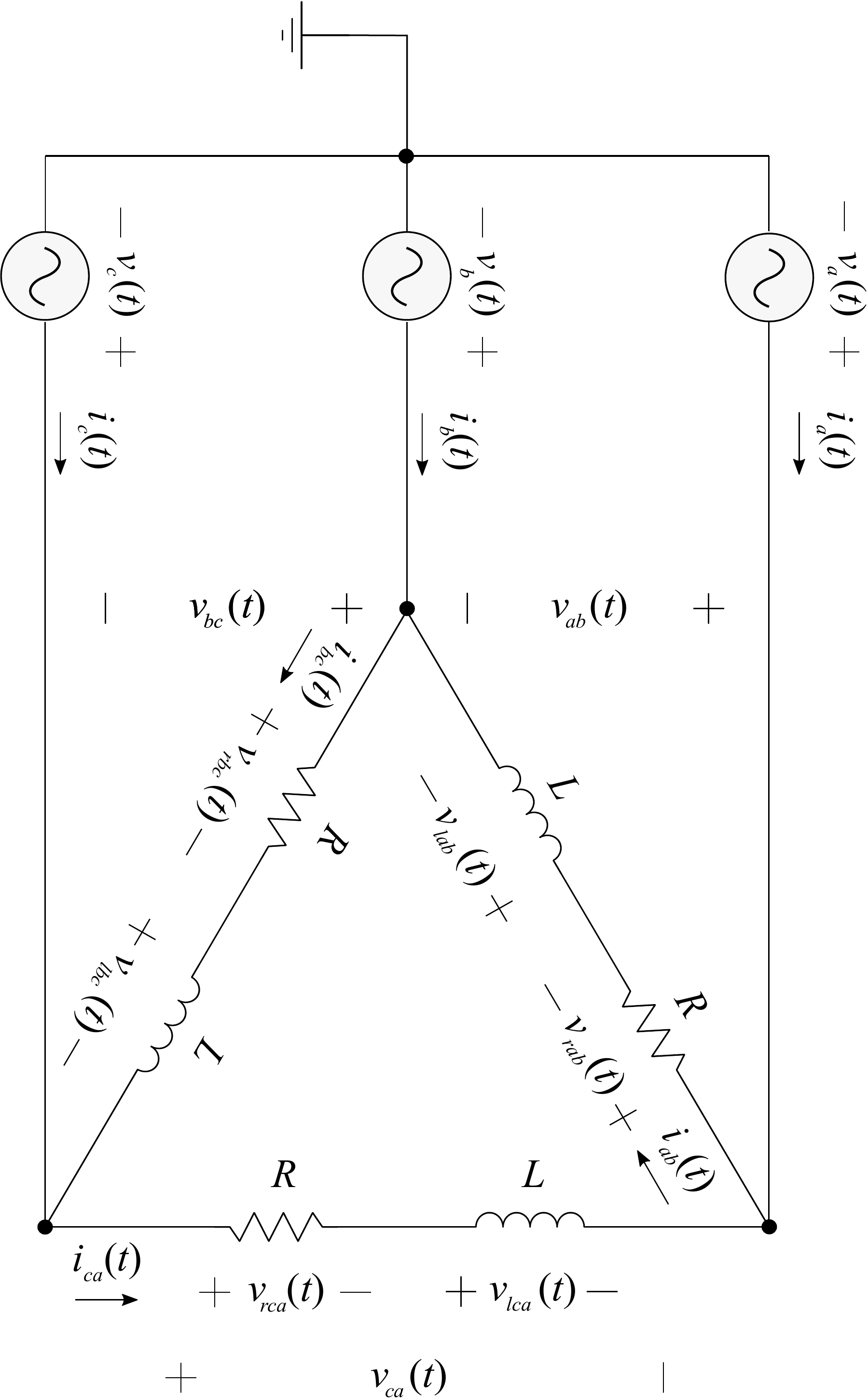}
\label{fig:delta-no-fault-dynamic}}
\hfil
\subfloat[Delta-connected RL load with a line-line fault]{\includegraphics[width=2.0in]{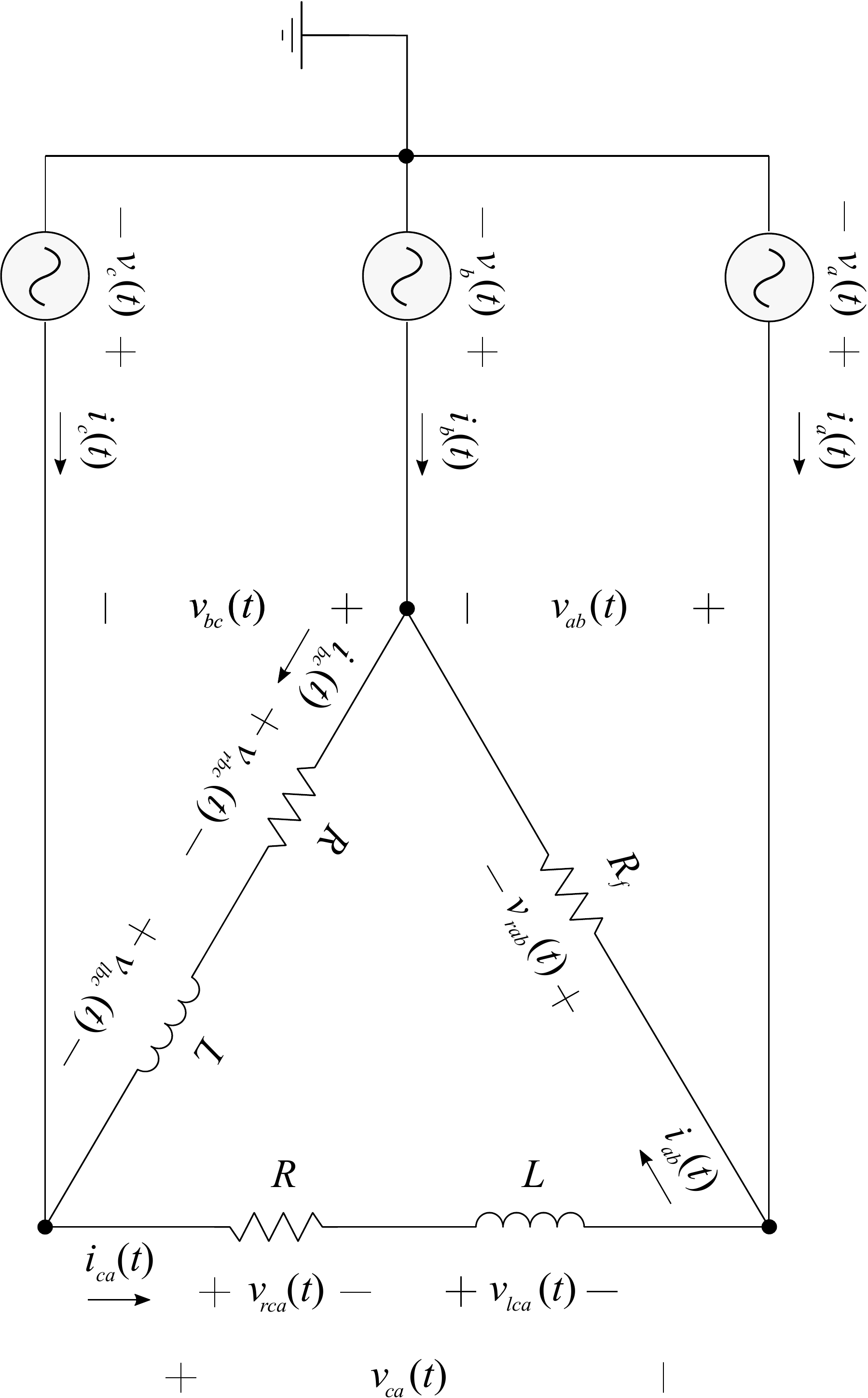}
\label{fig:delta-ll-fault-r-only-dynamic}}
\hfil
\subfloat[Delta-connected RL load with a line-ground fault]{\includegraphics[width=2.1in]{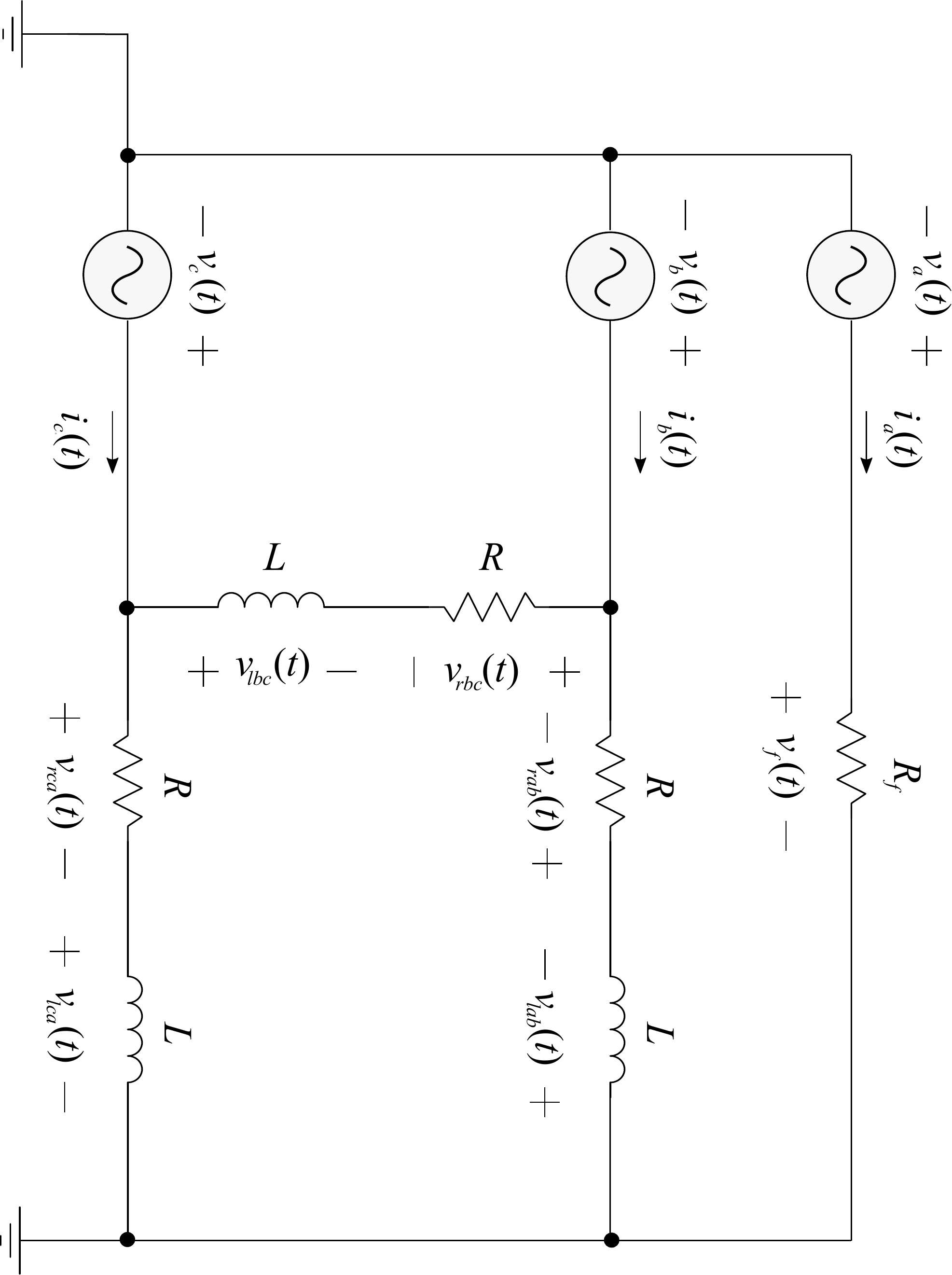}
\label{fig:delta-lg-fault-r-only-dynamic}}
\vspace{0.1in}
\caption{Dynamic implementation of state estimation-based protection, applied to single-phase, grounded-wye and delta-connected load configurations.}
\label{fig:dynamic-models}
\end{figure*}

Modeling assumptions throughout this paper are the same as those in \cite{barnes21-dse}.
Instead of ideal power sources, non-ideal sources (inverter model introduced in \cite{barnes2021implementing}) are used.
The operation of DSE for load fault detection is to apply a set of parallel state estimators, one for normal operation and one for each fault configuration. On a three-phase load, for example, the following state estimators exist: 1) Unfaulted; 2) Phase~A line-ground fault; 3) Phase~B line-ground fault; 4) Phase~C line-ground fault; 5) Phases~A-B line-line fault; 6) Phases~B-C line-line fault; 7) Phases~C-A line-line fault; 8) Three-phase fault.
The state estimators run in parallel and the current state of the system is indicated by the state estimator with the lowest error.
In this work, only the unfaulted, line-ground, and line-line cases are considered, as illustrated in Fig.~\ref{fig:dynamic-models}.

The problem formulation for the simplified grounded-wye-connected RL load with a line-line fault of Fig.~\ref{fig:gwye-ll-fault-r-only-dynamic} is described below.
\textit{For other load configurations, refer to earlier work \cite{barnes21-dse}.}

\vspace{0.1in}
\noindent
\normalsize The terminal equations for the case:
\small
\begin{subequations} 
\begin{align}
v_{ab}(t) = v_f(t) \qquad \quad
v_c(t) = v_r(t) + v_l(t)
\end{align}
\begin{align*}
i_a(t) = G_f v_f(t) \qquad \quad
i_b(t) = -G_f v_f(t)
\end{align*}
\begin{align}
i_c(t) = Gv_r(t) = \frac{1}{L} \int_0^t v_l(t) \; dt
\end{align}
\end{subequations}

\noindent
\normalsize The state equations for the case:
\small
\begin{subequations} 
\begin{align}
0 &= Gv_r(t) = \frac{1}{L} \int_0^t v_l(t) \; dt \\
0 &= G(v_r(t) - v_r(t - \Delta t)) = \Gamma \int_{t-\Delta t}^t v_l(t) \; dt \\
0 &= G(v_r(t) - v_r(t - 2\Delta t)) = \Gamma \int_{t-2\Delta t}^t v_l(t) \; dt    
\end{align}
\end{subequations}

\noindent
\normalsize The output and state of the system, respectively:
\small
\begin{subequations}
\begin{align}
y &=    \begin{bmatrix}
        v_{ab}(t) & v_c(t) & i_a(t) & i_b(t) & i_c(t) & z_c(t) 
        \end{bmatrix}^T \\
x &=    \begin{bmatrix}
        G & \Gamma & G_f & v_f(t) & v_r(t) & v_l(t)
        \end{bmatrix}^T
\end{align}
\end{subequations}

\noindent
\normalsize Assuming that the signals are sampled at time points $n \in \{1,\ldots,N\}$ with sample time $\Delta t$, the discrete-time state-output mapping function $h(x)$ therefore:

\small
\noindent $\forall n \in \{1,\ldots,N\}$
\begin{align}
\begin{split}
h_n(x) &= v_{ab}(n) = v_f(n) \\
h_{n+N}(x) &= v_c(n) = v_r(n) + v_l(n) \\
h_{n+2N}(x) &= i_a(n) = G_f v_f(n) \\
h_{n+3N}(x) &= i_b(n) = -G_f v_f(n) \\
h_{n+4N}(x) &= i_c(n) = G v_r(n) \\
\end{split}
\end{align}

\small
\noindent $\forall n \in \{3,\ldots,N\}$
\begin{align}
\begin{split}
h_{n+5N-2}(x) &= G(v_r(n) - v_r(n-2) \\
    & - \frac{2\Delta t\Gamma}{6}(v_l(n) + 4v_l(n-1) + v_l(n-2))
\end{split}
\end{align}

\noindent
\normalsize where the last equation comes from Simpson's $1/3$ rule \cite{chapra2009}:
\small
\begin{equation}
\int_{t-2\Delta t}^t f(t) \; dt = \frac{2\Delta t}{6}(f(t) + 4f(t-\Delta t) + f(t-2\Delta t)    
\end{equation}

\noindent
\normalsize The Jacobian of the system $\mathbf{H}$ therefore:

\small
\noindent $\forall n \in \{1, \ldots, N\}$
\begin{align}
\begin{split}
\frac{\partial v_{ab}(n)}{\partial v_f(n)} &= 1 \qquad
\frac{\partial v_c(n)}{\partial v_r(n)} = 1 \qquad
\frac{\partial v_c(n)}{\partial v_l(n)} = 1 \\
\frac{\partial i_a(n)}{\partial G_f} &= v_f(n) \qquad 
\frac{\partial i_a(n)}{\partial v_f(n)} = G_f \qquad
\frac{\partial i_b(n)}{\partial G_f} = -v_f(n) \\
\frac{\partial i_b(n)}{\partial v_f(n)} &= -G_f \qquad
\frac{\partial i_c(n)}{\partial G} = v_r(n) \qquad
\frac{\partial i_c(n)}{\partial v_r(n)} = G \\
\frac{\partial z(n)}{\partial v_l(n)} &= \frac{2\Delta t\Gamma}{6}
\end{split}
\end{align}

\vspace{0.05in}
\noindent
\textit{Note that for simplicity of notation, the indexing of individual elements of $\mathbf{H}$ is not presented.}

\vspace{0.1in}
\small
\noindent $\forall n \in \{2, \ldots, N\}$
\begin{align}
\frac{\partial z(n)}{\partial v_l(n-1)} = \frac{8\Delta t\Gamma}{6}
\end{align}

\vspace{0.1in}
\small
\noindent $\forall n \in \{3, \ldots, N\}$
\begin{align}
\begin{split}
\frac{\partial z(n)}{\partial \Gamma} = - \frac{2\Delta t\Gamma}{6}(v_l(n) + 4v_l(n-1) + v_l(n-2)) \\
\frac{\partial z(n)}{\partial v_r(n)} = G \qquad
\frac{\partial z(n)}{\partial v_r(n-2)} = -G \qquad 
\frac{\partial z(n)}{\partial v_l(n-2)} = \frac{2\Delta t\Gamma}{6}
\end{split}
\end{align}

\vspace{0.1in}
\normalsize The simplified model for a delta-connected load with line ground fault, illustrated in Fig.~\ref{fig:dynamic-models}g, is described as follows.
\vspace{0.05in}

\noindent
\normalsize The terminal equations of the system:
\small
\begin{subequations} 
\begin{align*}
v_a(t) = v_f(t) \qquad \quad
v_b(t) = -(v_{rab}(t) + v_{lab}(t))
\end{align*}
\begin{align}
v_c(t) = v_{rca}(t) + v_{lca}(t)
\end{align}
\begin{align*}
i_a(t) = G_f v_f(t) \qquad \quad
i_b(t) = G(v_{rbc}(t) - v_{rab}(t))
\end{align*}
\begin{align}
i_c(t) = G(v_{rca}(t) - v_{rbc}(t))
\end{align}
\end{subequations}

\noindent
\normalsize The state equations of the system:
\small
\begin{subequations}
\begin{align}
0 & = G v_{rab}(t) = \frac{1}{L}\int_0^t v_{lab}(t) \; dt \\
0 & = G v_{rbc}(t) = \frac{1}{L}\int_0^t v_{lbc}(t) \; dt\\
0 & = G v_{rca}(t) = \frac{1}{L}\int_0^t v_{lca}(t) \; dt
\end{align}
\end{subequations}

\noindent
\normalsize The output and state of the system, respectively:
\small
\begin{subequations}
\begin{align}
y &=    \left[\begin{smallmatrix}
        v_a(t) & v_b(t) & v_c(t) & i_a(t) & i_b(t) & i_c(t) & z_{ab}(t) & z_{bc}(t) & z_{ca}(t)
        \end{smallmatrix}\right]^T \\
x &=    \left[\begin{smallmatrix}
        G & \Gamma & G_f & v_{rab}(t) & v_{rbc}(t) & v_{rca}(t) & v_{lab}(t) & v_{lbc}(t) & v_{lca}(t) & v_f(t)
        \end{smallmatrix}\right]^T
\end{align}
\end{subequations}

\noindent
\normalsize The discrete-time state-output mapping function therefore:

\small
\noindent $\forall n \in \{1,\ldots,N\}$
\begin{align}
\begin{split}
h(n) &= v_f(n) \\
h(n+N) &= - v_{rab}(n) - v_{lab}(n) \\
h(n+2N) &= v_{rca}(n) + v_{lca}(n) \\
h(n+3N) &= G_f v_f(n) \\
h(n+4N) &= G(-v_{rab}(n) + v_{rbc}(n)) \\
h(n+5N) &= G(-v_{rbc}(n) + v_{rca}(n)) 
\end{split}
\end{align}

\small
\noindent $\forall n \in \{3,\ldots,N\}$
\begin{align}
\begin{split}
h(n+6N-2) &=  G(v_{rab}(n) - v_{rab}(n-2) \\
        & - \frac{2\Delta t\Gamma}{6}(v_{lab}(n) + 4v_{lab}(n-1) + v_{lab}(n-2)) \\
h(n+7N-2) &= G(v_{rbc}(n) - v_{rbc}(n-2) \\
        & - \frac{2\Delta t\Gamma}{6}(v_{lbc}(n) + 4v_{lbc}(n-1) + v_{lbc}(n-2)) \\
h(n+8N-2) &=  G(v_{rca}(n) - v_{rca}(n-2) \\
        & - \frac{2\Delta t\Gamma}{6}(v_{lca}(n) + 4v_{lca}(n-1) + v_{lca}(n-2)) 
\end{split}
\end{align}


\begin{figure*}[!htbp]
\centering

\subfloat[Grounded-wye-connected load]{\includegraphics[width=0.475\textwidth,angle=0]{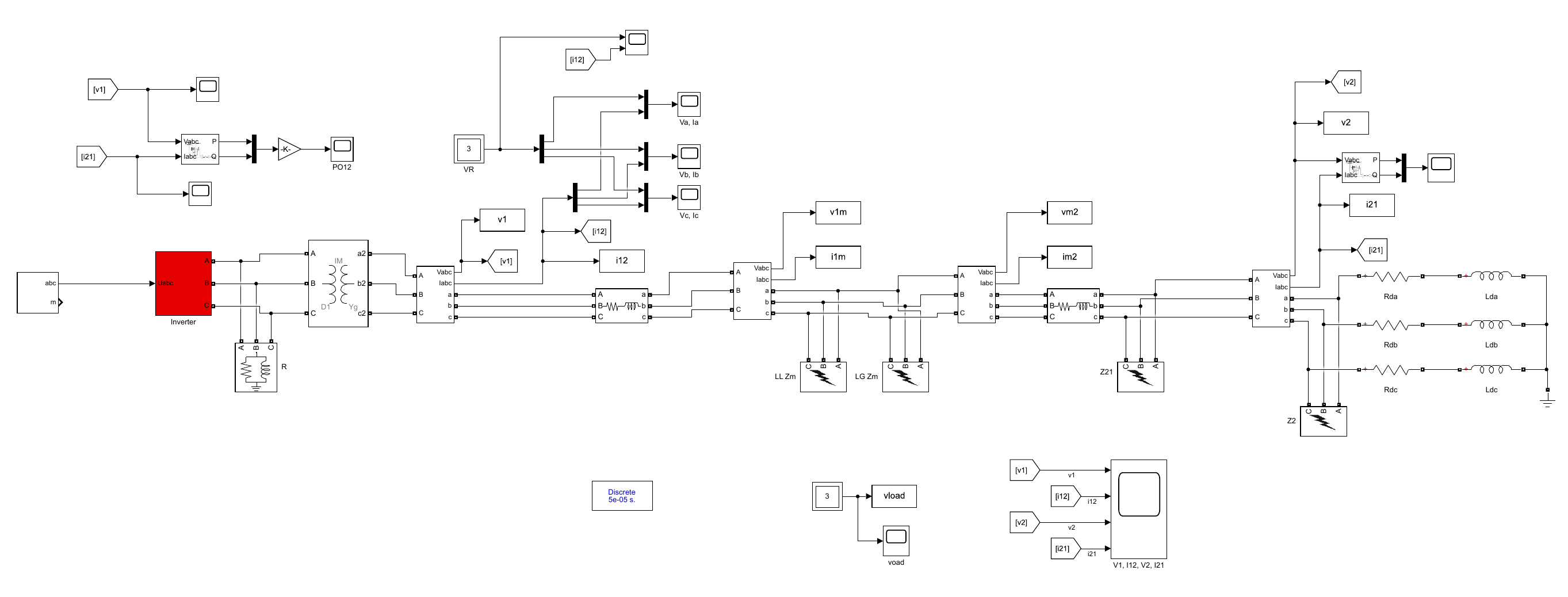}
\label{fig:simulink-gwye-microgrid}}
\hfil
\subfloat[Delta-connected load]{\includegraphics[width=0.475\textwidth,angle=0]{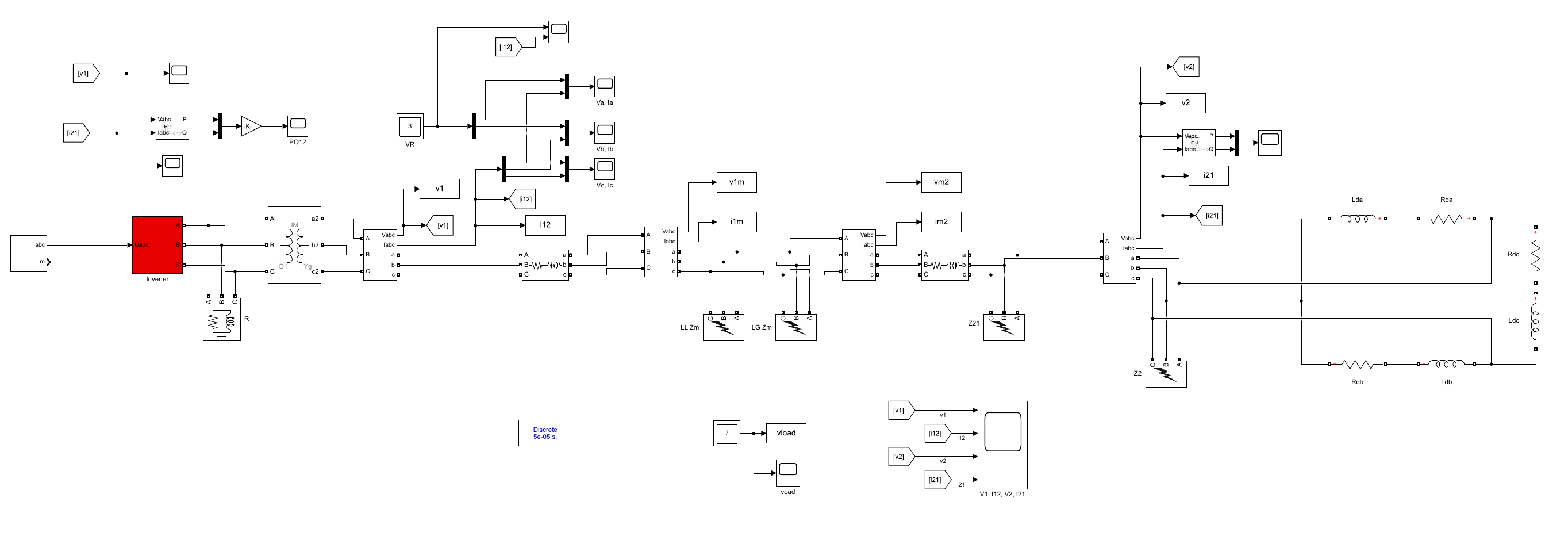}
\label{fig:simulink-delta-microgrid}}

\vspace{0.1in}
\caption{Simulink model of a two-bus microgrid with different loads \cite{vasquez_modeling_2013}.}
\label{fig:simulink-models}
\end{figure*}




\noindent
\normalsize The Jacobian of the system therefore:

\small
\noindent $\forall n \in \{1, \ldots, N\}$
\begin{align}
\begin{split}
\frac{\partial v_a(n)}{v_f(n)} &= 1 \qquad
\frac{\partial v_b(n)}{v_{rab}(n)} = -1 \qquad
\frac{\partial v_b(n)}{v_{lab}(n)} = -1 \\
\frac{\partial v_c(n)}{v_{rca}(n)} &= 1 \qquad
\frac{\partial v_c(n)}{v_{lca}(n)} = 1 \qquad
\frac{\partial i_a(n)}{R_f} = v_f(n) \\        
\frac{\partial i_a(n)}{v_f(n)} &= G_f \qquad
\frac{\partial i_b(n)}{v_{rbc}(n)} = G \qquad
\frac{\partial i_b(n)}{v_{rab}(n)} = -G \\
\frac{\partial i_c(n)}{v_{rca}(n)} &= G \qquad
\frac{\partial i_c(n)}{v_{rbc}(n)} = -G  
\end{split}
\end{align}

\vspace{0.1in}
\small
\noindent $\forall n \in \{3, \ldots, N\}$
\begin{align}
\begin{split}
\frac{\partial Z_{ab}(n-2)}{G} &= v_{rab}(n) - v_{rab}(n-2) \\
\frac{\partial Z_{bc}(n-2)}{G} &= v_{rbc}(n) - v_{rbc}(n-2) \\
\frac{\partial Z_{ca}(n-2)}{G} &= v_{rca}(n) - v_{rca}(n-2) \\
\frac{\partial Z_{ab}(n-2)}{v_{rab}(n)} &= G \qquad \quad
\frac{\partial Z_{bc}(n-2)}{v_{rbc}(n)} = G \\
\frac{\partial Z_{ca}(n-2)}{v_{rca}(n)} &= G \qquad \quad
\frac{\partial Z_{ab}(n-2)}{v_{rab}(n-2n)} = -G \\
\frac{\partial Z_{bc}(n-2)}{v_{rbc}(n-2)} &= -G \qquad \quad
\frac{\partial Z_{ca}(n-2)}{v_{rca}(n-2)} = -G
\end{split}
\end{align}

\vspace{0.1in}
\small
\begin{align}
\begin{split}
\frac{\partial z_{ab}(n-2)}{\partial \Gamma} &= - \frac{2\Delta t\Gamma}{6}(v_{lab}(n) + 4v_{lab}(n-1) + v_{lab}(n-2)) \\
\frac{\partial z_{bc}(n-2)}{\partial \Gamma} &= - \frac{2\Delta t\Gamma}{6}(v_{lbc}(n) + 4v_{lbc}(n-1) + v_{lbc}(n-2)) \\
\frac{\partial z_{ca}(n-2)}{\partial \Gamma} &= - \frac{2\Delta t\Gamma}{6}(v_{lca}(n) + 4v_{lca}(n-1) + v_{lca}(n-2)) \\        
\frac{\partial z_{ab}(n-2)}{\partial v_{lab}(n)} &= -\frac{2\Delta t}{6\Gamma} \qquad \quad
\frac{\partial z_{bc}(n-2)}{\partial v_{lbc}(n)} = -\frac{2\Delta t}{6\Gamma} \\
\frac{\partial z_{ca}(n-2)}{\partial v_{lca}(n)} &= -\frac{2\Delta t}{6\Gamma} \qquad \quad
\frac{\partial z_{ab}(n-2)}{\partial v_{lab}(n-1)} = -\frac{2 4\Delta t}{6\Gamma} \\
\frac{\partial z_{bc}(n-2)}{\partial v_{lbc}(n-1)} &= -\frac{2 4\Delta t}{6\Gamma} \qquad \quad
\frac{\partial z_{ca}(n-2)}{\partial v_{lca}(n-1)} = -\frac{2 4\Delta t}{6\Gamma} \\
\frac{\partial z_{ab}(n-2)}{\partial v_{lab}(n-2)} &= -\frac{2\Delta t}{6\Gamma} \qquad \quad
\frac{\partial z_{bc}(n-2)}{\partial v_{lbc}(n-2)} = -\frac{2\Delta t}{6\Gamma} \\
\frac{\partial z_{ca}(n-2)}{\partial v_{lca}(n-2)} &= -\frac{2\Delta t}{6\Gamma} \\
\end{split}
\end{align}

\vspace{0.1in}
\normalsize Given $h(n)$ and $H(n,n)$, the state of the system can be solved by the following updated equations:

\small
\begin{subequations}
\begin{align}
\epsilon_i &  = y - h(x_i) \\
x_{i+1} & = x_{i} + (\mathbf{H}_i^T \mathbf{H}_i)^{-1}\mathbf{H}_i^T\epsilon_i
\end{align}
\end{subequations}

\noindent
\normalsize This process is repeated iteratively until either the maximum number of iterations is reached or the algorithm has converged, indicated by the change in the log of the squared error falling below a specified threshold:

\small
\begin{equation}
J_i = \log |\epsilon_i^* \epsilon_i|
\end{equation}

\newpage
\normalsize

\section{Dynamic Grid-Forming Inverter Model \\ with Current-Limiting} \label{sec:inverter-model-with-current-limiting}
\indent


The algorithm described in Section~\ref{sec:methodology} was tested on a four-bus case study system, introduced in \cite{barnes2021implementing} and illustrated in Fig.~\ref{fig:simulink-models}.
Two versions of the case study system were developed: one that includes a grounded-wye connected load of the original system (Fig.~\ref{fig:simulink-gwye-microgrid}) and one that includes a delta-connected load of the original system (Fig.~\ref{fig:simulink-delta-microgrid}). Both of these make use of the grid-forming inverter model with proportional-resonant control (described in \cite{vasquez_modeling_2013}), which was selected as it provides good voltage regulation under unbalanced or nonlinear loads compared with traditional delta-quadrature control.

Previous work, presented in \cite{barnes21-dse}, has demonstrated that the behavior of grid-forming inverters is strongly affected by the choice of current-limiting strategy.
This paper employs a hysteresis strategy (described in \cite{bottrell_comparison_2014}): when an overcurrent condition is detected, the inner current-control loop of the inverter is switched from the output of the outer voltage-control loop to a fixed current reference; when the controller determines that the fault has been removed from the system, it will switch back to normal operation. This detail of control, however, is not fully modeled here; the hysteresis approach is selected as compared to a naive current-limiting approach, it does not result in the injection of voltage harmonics into the system that could impair the ability of certain protection methods to operate.

\begin{figure}[!htbp]
\centering
\includegraphics[width=0.45\textwidth]{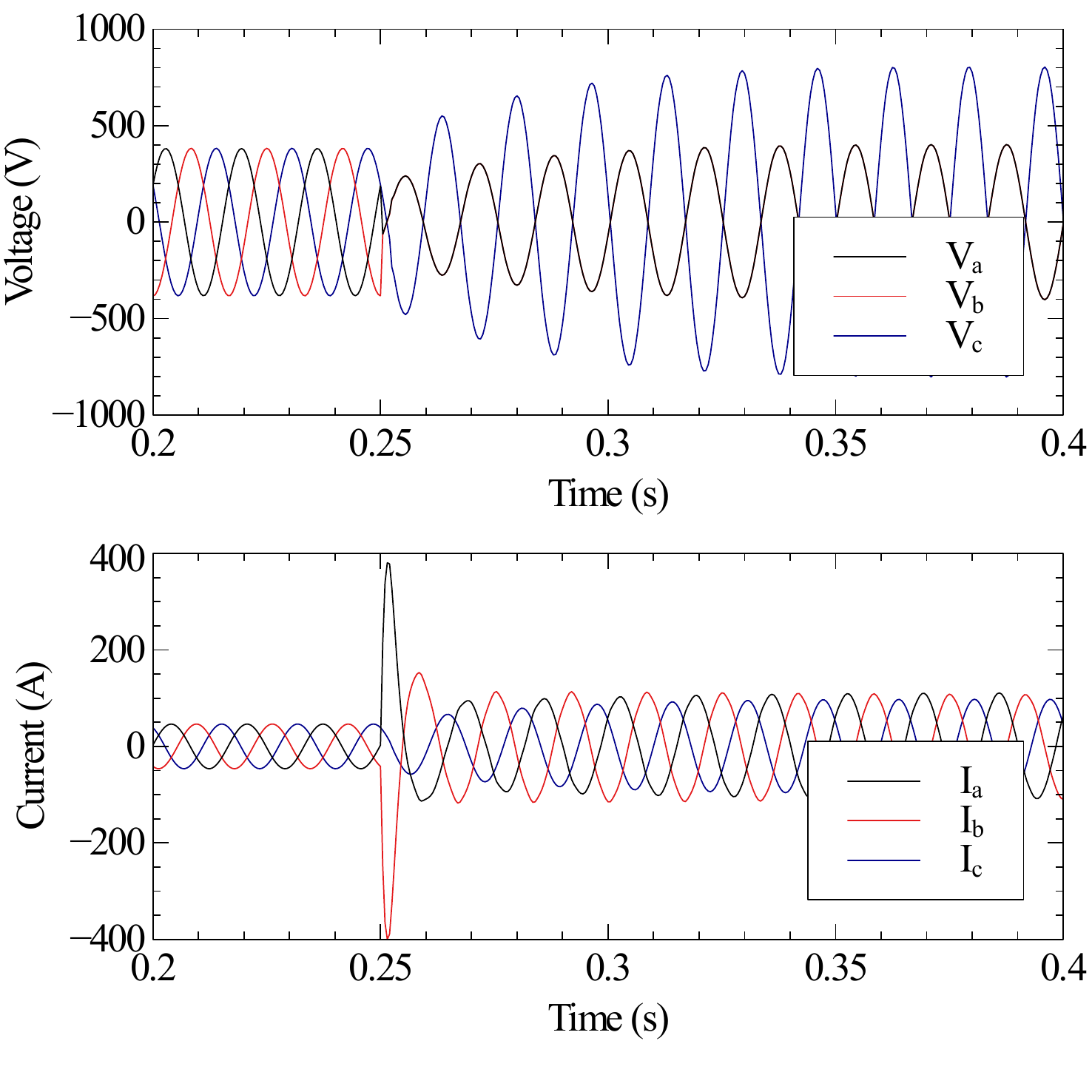}
\caption{Voltage and current at load bus during a line-ground load fault for a grounded-wye load configuration.}
\label{fig:gwye-lgf-vi}
\end{figure}

For each case study system, both line-ground and line-line faults are applied.
Fig.~\ref{fig:gwye-lgf-vi} illustrates the transient behavior of the inverter during a line-ground fault when it enters into current-limiting mode at 250~ms. To avoid handling the transient behavior of the system during fault inception, the state-estimation algorithm is applied starting at 300~ms, after the initial transients have died down.

\begin{table*}[!htbp]
\caption{Results for Dynamic State Estimation}
\begin{center}
\begin{tabular}{lrrrrrrr}
\hline
Case & $R$ ($\Omega$) & $\hat{R}$ ($\Omega$) & $L$ (mH) & $\hat{L}$ (mH) & $R_f$ ($m\Omega$) & $\hat{R}_f$ ($m\Omega$) \\ 
\hline \hline
Single-Phase RL Load & 7.373 & 7.362  & 9.779 & 9.841 & 15.000 & 12.272 \\
Grounded-Wye No Fault & 7.373 & 7.369 & 9.779  & 9.784 & -- & -- \\
Grounded-Wye Line-Ground Fault & 7.373 & 7.373 & 9.779 & 9.779 & 15.000 & 14.976 \\
Grounded-Wye Line-Line Fault & 7.373 & 7.373 & 9.779 & 9.779 & 10.000 & 9.995 \\
Delta No Fault & 7.373 & 7.368 & 9.779 & 9.800 & -- & -- \\
Delta Line-Line Fault  & 7.373 & 10.150 & 9.779 & 16.981 & 10.000 & 9.922 \\
Delta Line-Ground Fault & 7.373  & 0.00 & 9.779 & 45.460 & 15.000 & 14.580 \\
\hline
\end{tabular}
\end{center}
\label{table:dynamic-results}
\end{table*}


\section{Results} \label{sec:results}
\indent

The results of the state estimation algorithm are displayed in Table~\ref{table:dynamic-results}.
Previous work, presented in \cite{barnes21-dse}, observed convergence difficulties with a line-line fault on a grounded-wye connected load, presumably because the total number of parameters was high relative to the number of observables.
The simplified model, illustrated in Fig.~\ref{fig:gwye-ll-fault-r-only-dynamic}, helps overcome this issue for fault scenarios where $R_f \ll R + j\omega L$.
The fault models for delta-connected loads still have accuracy issues with estimating the load parameters on unfaulted phases, though they produce more accurate estimates of the fault resistance on the faulted phase(s).

\section{Conclusions} \label{sec:conclusion}
\indent


This paper has demonstrated that the DSE load-protection method presented for loads supplied by ideal voltage sources can be applied to loads on inverter-interfaced microgrids, and that for sufficiently low fault resistances approximate models can give effective convergence.

A number of issues remain to be addressed.
First, DSE for line protection requires measurements at both terminals of the line and high-speed communication between the terminals to provide time-series voltage measurements
Second, specific to load protection, to provide a sufficient ratio of observable to estimated parameters, load models described in this paper assume balanced conditions, which could potentially result in an erroneous trip under unbalanced loading conditions. This could also result in an an erroneous trip should a downstream single-phase protective device activate, causing the net load on the three-phase portion of the system to become unbalanced.


\newpage
\bibliographystyle{unsrt}
\bibliography{references}

\end{document}